**Deser formula for the shift of energy levels of hadronic atoms in terms of the effective radius of the strong interaction with the Coulomb wave function of hadronic atoms at the origin**

*O. Voskresenskaya*
*Joint Institute for Nuclear Research, Dubna*
*email: voskr@jinr.ru*

**Abstract.** It is shown that the relations between probabilities of $A_{2\pi}$-atoms creation in *ns*-states, derived with neglecting of strong interaction between pions, remain practically unchanged if the strong interaction is taken into account in the first order of perturbation theory. The formulation of Deser equation for the energy shift of energy levels of hadronic atoms is given in terms of the effective range of strong interaction and the relative correction to the Coulomb wave function of hadronic atoms at the origin, caused by the strong interaction.

## Introduction

The study of elementary atoms (hydrogen-like atoms formed by two oppositely charged elementary particles – not necessarily hadrons) has attracted attention from physicists, both theoretical and experimental [see, for example, 1–9]. The measurement of the shifts of levels of the hadronic atoms due to the strong interaction and their ground-state lifetimes enables to determine the values of hadron-hadron scattering lengths with an accuracy exceeding that achievable in direct low-energy hadron–hadron scattering experiments and thereby test the predictions of the QCD-inspired chiral symmetry model [4, 5]. Experiments to study the properties of $A_{\pi^- p}$, $A_{\pi^- d}$ [6]; $A_{K^- p}$ [7], $A_{K^- d}$ [8]; $A_{\pi^- \pi^+}$ and $A_{\pi^- K^+}$ [9] are currently being conducted or planned. The theoretical basis for the experimental determination of hadron–hadron scattering lengths by studying the properties of hadronic atoms is the Deser formula [10] (see also [11]), derived within the nonrelativistic quantum-mechanical approach

$$\Delta E_{ns}^{s} = -\frac{2\pi}{\mu} a_s |\psi_{ns}^{c}(0)|^2, \tag{1}$$

$$\Gamma_{1s} = \tau_{1s}^{-1} = \frac{16\pi}{9}\left(\frac{p^*}{\mu}\right) |a_{ce}|^2 |\psi_{1s}^{c}(0)|^2. \tag{2}$$

Here $\mu$ is the reduced mass of the hadron atom; $p^*$ is the momentum of the products of the annihilation of the atom into a pair of neutral hadrons; $\psi_{ns}^{c}(0)$ is the value of the Coulomb wave function of the hadron atom for coinciding values of the radius vectors of its constituent hadrons.

The scattering lengths $a_s$ and the charge exchange lengths $a_{ce}$ are related to scattering lengths in certain isotopic states of the hadron–hadron system $a_I$ by the relations: $a_s = (2a_{1/2} + a_{3/2})/3$, $a_{ce} = (a_{1/2} + a_{3/2})$ for $\pi^- p$-, $\pi^- K$-atoms and $a_s = (2a_0 + a_2)/3$, $a_{ce} = (a_0 + a_2)$ for $\pi^+ \pi^-$- atoms.

A characteristic property of equations (1) and (2) is the factorization of the contributions of the strong (represented by scattering lengths) and electromagnetic (represented by Coulomb wave functions) interactions, which is a consequence of the approximations used in their derivation.

Numerous studies are devoted to assessing the accuracy of relations (1), (2) and calculating relativistic, electromagnetic and other corrections to them within various field-theoretical approaches [12], e.q. the non-relativistic approach of effective Lagrangians [13] and chiral perturbative theory [14], based on the formalism of the Bethe-Salpeter equations [15]). The corrections to the formula for the lifetime of pionium atoms in the ground state, the experimental measurement of which is being carried out at CERN by the DIRAC collaboration [1–3, 9], have been studied in most detail.

It is shown that for the main pionium decay channel $\Gamma = \Gamma^{2\pi^0} + \Gamma^{2\gamma} + \ldots \approx \Gamma^{2\pi^0} (\Gamma^{2\gamma} \approx 0{,}36\Gamma^{2\pi^0})$, the relativistic correction to the decay width $\Gamma_o = \Gamma_{1s}$ in the first order of perturbation theory is 0.5570, the electromagnetic correction is 1.770 and the total correction, calculated taking into account all possible effects, is on the order of $\sum_i \delta_i = (6{,}1 \pm 3.1)\% \approx \delta_\Gamma = (5{,}8 \pm 1{,}2)\%$ [15]. Taking these effects into account, the decay width from the ground state is estimated by

$$\Gamma = \Gamma_0(1 + \delta_\Gamma). \tag{3}$$

According to estimates in [1–3, 9], the theoretical uncertainty in the expression relating the lifetime of pionium atoms to the $\pi$-scattering lengths is of the order of 1% [9]. This uncertainty would result in only of a 0.5% error in determining the combination of $a_0 - a_2$ scattering lengths if the DIRAC experiment could measure the lifetime of pionium atoms with absolute precision, which is, surely, impossible.

The method for determining the lifetime of pionium atoms in this experiment is based on a comparison of the experimentally measured ionization probability of pionium atoms with the theoretical calculation of this value, which includes $T_{1S}$ as an adjustable parameter. Therefore, the accuracy of the measurement is defined not only by the errors of the experiment itself, but also by the accuracy of the approximations used in the theoretical description of the processes of pionium atom formation in the target and their subsequent passage through the substance.

Below, we discuss the accuracy of the fundamental relations of dimesoatom formation theory [16], used in interpreting the DIRAC experiment. A key role in this theory is played by the relation

$$\left|\frac{\psi_{n_1 s}(r)}{\psi_{n_2 s}(r)}\right|^2_{r\to 0} = \left(\frac{n_2}{n_1}\right)^3 + O\left(\frac{<r^2>_p}{r_B^2}\right) \tag{4}$$

between the values of the squares of the wave functions in $\pi$-states at small values ($\tau \ll \tau_B$) of the argument ($\tau_B$ is the Bohr radius of the atom).

This relation is strict if only Coulomb forces act between the particles that make up the dimesoatom, and could, in principle, be violated if the strong interaction between them is taken into account.

Further, in first-order perturbation theory in the intensity of the strong interaction, we will show that this relation remains valid with an accuracy of $(r_s/r_B) \sim 10^{-3}$ even when the strong interaction between mesons is taken into account ($r_s$ is the characteristic radius of strong interactions). We will also reformulate the Deser formula (1), in terms of the effective radius of the strong interaction and the relative correction to the Coulomb wave function of pionium at zero $c_n = \triangle\,\psi_{ns}(0)/\psi^c_{ns}(0)$ due to the strong interaction, and show that, along with the s-scattering wave lengths $a_{I_1}$ and $a_{I_2}$, the quantities $c_n$ and $r_s$ could be determined experimentally.

## 1 Elements of the theory of the formation of elementary hadronic atoms

The problem of the formation of elementary hadronic atoms in high-energy hadron-hadron (or hadron-nuclear) interactions was first considered in the work of L.L. Nemenov [16], which essentially paved the way for systematic research in physics of relativistic hadronic atoms and laid the theoretical foundation for the experiment underway at CERN to measure the lifetime of the $\pi^+\pi^-$ atom (2) with an accuracy of about 10% [9]. Therefore, in further analysis, we will primarily follow the logic of this work.

Leaving aside physical details unimportant for further consideration, the main result of [16] for the probability of the formation of a pionium atom in (2) a certain state $f$ ($f$ is the set of quantum numbers describing this state) can be represented as

$$\mathfrak{w}_f = \left|\int M(\vec{p})\psi_f(\vec{p})d^3p\right|^2. \quad (5)$$

Here, $\mathfrak{w}_f(\vec{p})$ is the wave function of the pionium atom in the state $f$, which is the Fourier transform

$$\psi_f(\vec{p}) = \frac{1}{(2\pi)^{3/2}}\int \psi_f(\vec{r})e^{i\vec{p}\vec{r}}d^3r \quad (6)$$

of the wave function of the pionium atom in the coordinate representation; $M(\vec{p})$ is the amplitude of the formation of a $\pi+\pi-$ pair with the relative momentum $\vec{p}$ during the interaction of the proton beam with the target, normalized by the condition

$$\sum_f \mathfrak{w}_f = \int |M(\vec{p})|^2 d^3p = 1. \quad (7)$$

The wave functions $\psi_f(\vec{r})$ are solutions of the Schrödinger equation

$$\left[-\frac{\Delta}{2\mu} + V(r)\right]\psi_f(\vec{r}) = E_f\psi_f(\vec{r}), \quad (8)$$

where $\mu = m_1m_2/(m_1+m_2)$ is the reduced mass, V(r) is the sum of the potential energies of the electromagnetic (Coulomb) and strong interactions

$$V(r) = V_c(r) + V_s(r)\,. \quad (9)$$

The form of the Coulomb potential is well known, i.e.

$$V_c(r) = -\alpha/r, \quad \alpha = 1/137, \quad r = |\vec{r}|\,.$$

The form of the strong interaction potential could be reconstructed using inverse scattering methods in the presence of comprehensive experimental information on the energy dependence of the phase shifts of elastic $\pi+\pi-$ scattering in the $s$-state. Here and below, the spectroscopic symbol *s* is used to denote quantum-mechanical states with zero values of the orbital (*l)* and, accordingly, magnetic ($\tau$) quantum numbers. However, the volume of this information is very limited and its accuracy, as a rule, is low, which has prevented the reconstruction of the $V_S(r)$ function. Therefore, to estimate the order of magnitude of various effects due to the strong $\pi+\pi-$ interaction, expressions for $V_S(r)$ obtained within certain theoretical models are usually used.

Since different models yield quite different expressions for $V_S(r)$, a significant model dependence of the results arises, leading to uncertainties in theoretical predictions at the level of several percent, and sometimes tens of percent.

As will be shown below, uncertainties of precisely this order appear when calculating the absolute values of probabilities (5). However, in relations of the form $\mathfrak{w}_{f_1}/\mathfrak{w}_{f_2}$ (relative probabilities), they are significantly reduced, so that the accuracy of calculating the latter turns out to be quite high.

Before discussing the influence of the strong π+π- interaction on the quantities (5) and the relative probabilities, we will present the results of considering this problem, obtained by neglecting it.

### 1.1 Formation of Coulomb dimesoatoms by a point source

If the strong interaction between pions were absent or, for some reason, had only a weak effect on the values of the wave functions of the pionium atom [17, 18], so that the value of $V_S(r)$ in the Schrödinger equation could be neglected, then the wave functions of the pionium atom would be solutions of the quantum-mechanical Keplerian problem and, up to a scaling transformation, would coincide with the wave functions of the hydrogen atom [19]. For simplicity, pionium atoms described by such wave functions will be called Coulomb pionium atoms. The distribution of relative momentum in Coulomb atoms is concentrated in a narrow region $\vec{p}\sim p_B = \mu\alpha$, where $\mu$ is the reduced mass, equal in this case to half the pion mass $\mu = m_\pi/2 \approx 70$ MeV, so that $p_B\sim \approx 0.5$ MeV (the system of units $\hbar$ = c = 1 is adopted throughout this article).

Since the quantity M($\vec{p}$) changes much more slowly than ψ($\vec{p}$) (it is characterized by the momentum $|\vec{p}|{\sim}p_s{\sim}100$ MeV), in the approximation used, it can be taken out from under the integral sign at the point $\vec{p}$ = 0, resulting in

$$\mathfrak{w}_f = \left|M(p=0)\int \psi_f(\vec{p})d^3p\right|^2 = (2\pi)^3\left|M(p=0)\int \psi_f(\vec{r}=0)\right|^2 . \quad (10)$$

It follows that pionium atoms can form in strong interactions only in those states *f* whose wave functions do not vanish at $\vec{r}$ = 0, i.e., in s-states. These states differ from each other only in the value of the principal quantum number n.

But for the ns-states of Coulomb dimesoatoms, according to [16]

$$\left|\psi_f(\vec{r}=0)\right|^2 = \frac{1}{n^3}\frac{(\mu\alpha)^3}{\pi}, \quad (11)$$

from which we obtain

$$\frac{\mathfrak{w}_{n_1 s}}{\mathfrak{w}_{n_2 s}} = \left(\frac{n_2}{n_2}\right)^3 . \quad (12)$$

If we introduce the quantity

$$M(\vec{r}) = \frac{1}{(2\pi)^{3/2}}\int M(\vec{p})\, e^{-i\vec{p}\vec{r}}d^3p , \quad (13)$$

which can be interpreted as the amplitude of the formation of a π+π- pair with a relative distance $\vec{r}$, and rewrite (5) as

$$\mathfrak{w}_f = \left|\int M(\vec{r})\psi_f\,(\vec{r})d^3r\right|^2, \quad (14)$$

then using the approximation that led to result (12) is equivalent to the assumption that in strong interactions, π+π- pairs are formed with practically zero relative distance, i.e.

$$M(\vec{r}) = M(\vec{p}=0)\delta(\vec{r}) , \quad (15)$$

where $\delta(\vec{r})$ is the Dirac delta function.

### 1.2 Formation of Coulomb dimesoatoms by an extended source

In reality, the dimensions of the region in which π+π- pairs are formed are finite, although small compared to the sizes of the pionium atoms into which some of these pairs are transformed as a result of Coulomb attraction. According to various estimates [20], the root-mean-square radius of this region is of the order of $< \vec{r}^2 >^{1/2} \sim 3 \div 10\ + 1$ fm, while the Bohr radius of the ground state of pionium is $r_B$ = 367 fm. Therefore, corrections of the order of $r_p/r_B$, where $r_p \sim < \vec{r}^2 >^{1/2}$ to the zeroth-order approximation result can amount to several percent.

Indeed, it is easy to see [21] that, with an accuracy of up to values of the order of $r_p/r_B$, the result for the probability of the formation of a pionium atom in the ns-state can be written in the form $\mathfrak{w}_n$ = 1

$$\mathfrak{w}_n = |M(p=0)|^2\left|\psi_f(\vec{r}=0)\right|^2\left[1 + 2 < r >_p \frac{\psi_n'(r=0}{r\psi_n(r=0}\right], \quad (16)$$

where

$$< r >_p = \frac{\int M(\vec{r})rd^3\vec{r}}{\int M(\vec{r})d^3\vec{r}} \sim < r^2 >_p^{1/2} . \quad (17)$$

Since only the order of magnitude of the size of the region of the formation of π+π- pairs and, accordingly, the quantity $< r >_p$, is known, but not its exact value, this introduces uncertainty into the value of $\mathfrak{w}_n$.

But if we again limit ourselves to considering only the relative probabilities, which are the only ones that matter, then, given that the ratio

$$\frac{\psi_n'(r=0)}{\psi_n(r=0)} = \mu\alpha = \frac{1}{r_B} \quad (18)$$

does not depend on n, we obtain

$$\frac{\mathfrak{w}_{n_1 s}}{\mathfrak{w}_{n_2 s}} = \left(\frac{n_2}{n_1}\right)^3 + O\left(\frac{<r^2>_p}{r_B^2}\right), \quad (19)$$

where

$\left(\frac{<r^2>_p}{r_B^2}\right) \sim 10^{-3}$ .

Thus, we come to the conclusion that if the strong interaction between $\pi^+$ and $\pi^-$ in the pionium atom were absent, then result (12), obtained in [16], would be valid with a high degree of accuracy. This is ensured by the smallness of the quantity $< r^2 >_p / r_B^2$ and independence of the quantity from $n$.

The goal of subsequent analysis is to show that result (19) remains valid even when the effects of the strong $\pi^+\pi^-$-interaction are included in the consideration.

### 2.1 Estimation of the effect of the strong interaction on the wave functions of hadronic atoms in first-order Rayleih-Shrödinger perturbation theory

The ratios of the probabilities of the formation of hadronic atoms in various *ns*-states (12), (19) were obtained under the assumption that the wave functions of these states are purely Coulomb, i.e., satisfy the Schrödinger equation

$$\left[-\frac{\Delta}{2\mu} + V(r)\right]\psi(\vec{r}) = E\psi(\vec{r}) \,, \tag{20}$$

$$V(r) = V_c(r) \,.$$

Therefore, the influence of the strong interaction on both the values of the energy eigenvalues of bound (atomic) states of hadrons and on the spatial distribution of hadrons in atoms, described by their wave functions, was neglected. In what follows, we will remove this restriction and assume that the wave functions of hadronic atoms are solutions of the Schrödinger equation (20), in which the interaction potential energy, or more precisely, the potential *V(r)*, is the sum of the Coulomb potential and the strong interaction potential

$$V(r) = V_c(r) + V_s(r). \tag{21}$$

A specific property of the strong interaction is its short-range: the strong interaction potential is much larger than the Coulomb potential only in a very narrow range $r \leq r_s$ and becomes negligibly small compared to it in the range $r_s \ll r \ll r_B$.

For this reason, strong interaction effects are often corrective to Coulomb interaction effects. For example, they change the eigenvalues of the Hamiltonian operator of the problem under consideration very slightly (by no more than tenths of a percent), and this result is more or less model-independent. On this basis, one can attempt to treat them as perturbations and study them using well-known methods of perturbation theory.

Several variants of perturbation theory are known; their description can be found, for example, in [22]. The choice of a particular variant is determined by the specifics of the problem under study. The most general method of perturbation theory, called the "Rayleigh-Schrödinger method" and taught in all courses on quantum mechanics, is applicable to arbitrary quantum mechanical systems. Its essence relies on expanding the state vector (wave function) of the perturbed problem

$$H\psi_n = E_n\psi_n \,, \tag{22}$$

$$H = H_0 + V \,, \quad V = V_s \,, \quad H_0 = -\frac{\Delta}{2\mu} + V_c \,,$$

in a series over the complete system of state vectors $\psi_n^{(0)}$ of the unperturbed problem

$$H_0\psi_n^{(0)} = E_n\psi_n^{(0)} \,, \tag{23}$$

$$\psi_n = \psi_n^{(0)} + \Delta\psi_n$$

$$\Delta\psi_n = \sum_{k\neq n} c_{nk}\psi_k^{(0)} \tag{24}$$

In first-order perturbation theory, the coefficients of this expansion have the form

$$c_{nk}^{(1)} = \frac{V_{nk}}{E_n^{(0)} - E_k^{(0)}}, \tag{25}$$

Where $V_{nk} = \langle \psi_{\text{т}}^{(0)} | V | \psi_k^{(0)} \rangle$ is the matrix element for the states $\psi_n^{(0)}$ and $\psi_k^{(0)}$ of the unperturbed system of the perturbation operator $V$, and $E_n^{(0)}$ $E_k^{(0)}$ are the eigenvalues of the unperturbed Hamiltonian. In [17, 18], Rayleigh–Schrödinger perturbation theory was applied to estimate the influence of the strong meson–meson interaction on the wave functions of the ground states of $\pi^+\pi^-$ and $\pi^+K^-$-dimesoatoms at the origin.

In [17], the effective potential $V_S(r)$ of the $\pi^+\pi^-$ interaction was selected such that it reproduced the results of calculating the low-energy scattering amplitude within the virton–quark model.

The correction to the wave function of the ground state of $A_{2\pi}$- and $A_{\pi K}$-atoms was estimated in [17] using the relations

$$\Delta\psi_{1s}(0) = c_{12}\psi_{2s}^{(0)}(0), \tag{26}$$

$$c_{12}^{(1)} = \frac{\int \psi_{2s}^{(0)}(r) V_s(r) \psi_{1s}^{(0)}(r) d^3r}{E_{2s}^{(0)} - E_{1s}^{(0)}}. \tag{27}$$

Here $\psi_{1s}^{(0)}$ and $\psi_{2s}^{(0)}$ are the unperturbed (Coulomb) wave functions of the *1s*- and *2s*-states. The analysis was performed in a spherical basis, with the states indexed by the values of the principal (*n*), orbital (*l*), and magnetic (m) quantum numbers, so that, for example, $|2s\rangle \equiv |200\rangle$.

The result of this estimate for $A_{2\pi}$- and $A_{\pi K}$- atoms, respectively, is as follows:

$$\frac{\Delta\psi_{1s}(0)}{\psi_{1s}^{(0)}(0)} = 3{,}3 \cdot 10^{-4}, \quad \frac{\Delta\psi_{1s}(0)}{\psi_{1s}^{(0)}(0)} = 4.2 \cdot 10^{-4} \tag{28}$$

The authors of [18], relying on the results of the analysis of ππ- and πK-scattering within chiral models, approximated the potentials of $\pi^+\pi^-$- and $\pi^\pm K^\mp$-strong interactions by the expression

$$V_s(r) = g\delta(\vec{r}), \quad g = \frac{a_s}{4\pi}, \tag{29}$$

where $a_s$ is the length of $\pi^+\pi^-$- and $\pi^\pm K^\mp$-scattering “in the absence” of Coulomb interaction.

In this work, the correction to the value of the ground-state wave function ($|1s\rangle \equiv |100\rangle$) is estimated using the formulas

$$\Delta\psi_{1s} = \sum_{k=2}^{\infty} c_{1k}\psi_{ks}^{(0)}(0), \tag{30}$$

$$c_{1k}^{(1)} = \frac{\int \psi_{ks}^{(0)}(r) V_s(r) \psi_{1s}^{(0)}(r) d^3r}{E_{ks}^{(0)} - E_{1s}^{(0)}}, \tag{31}$$

$$\psi_{ns}(r) \equiv \psi_{n00}(r).$$

The results of calculations [18] for $A_{2\pi}$- and $A_{\pi K}$-atoms differ slightly from the estimates in [15]:

$$\frac{\Delta\psi_{1s}(0)}{\psi_{1s}^{(0)}(0)} = 2.5 \cdot 10^{-4}, \quad \frac{\Delta\psi_{1s}(0)}{\psi_{1s}^{(0)}(0)} = 1.1 \cdot 10^{-3}. \tag{32}$$

Thus, according to the estimates in [17, 18], the influence of the strong hadron-hadron interaction on the behavior of the wave functions of dimesoatoms at short distances is negligible, and when analyzing the probabilities of the formation of these atoms in the interaction of high-energy particles with nuclear targets, the wave functions of hadronic atoms can be considered purely Coulomb.

Qualitatively, the results in [17, 18] can be represented in the form

$$\frac{\Delta\psi(0)}{\psi^{(0)}(0)} \sim \frac{a_s}{r_B}, \tag{33}$$

where $a_s$ is the meson–meson scattering length, and $r_B$ is the Bohr radius of the corresponding dimesoatom.

However, it is shown in [23] that the calculations in the above-mentioned studies underestimate the influence of the strong interaction on the wave functions of $\pi^+\pi^-$- and $\pi^\pm K^\mp$-atoms

$$\Delta\psi_n^{(1)} = \sum_{k\neq n} c_{nk}^{(1)}\psi_k^{(0)}, \qquad c_{nk}^{(1)} = \frac{V_{nk}}{E_n^{(0)} - E_k^{(0)}}.$$

Returning to formulas (24), (25) for first-order perturbation theory, which determine the correction to the value of the wave function of the n-th state due to the perturbation potential $V_s$, we note that the sign of the sum implies not only summation over the states of the discrete spectrum, but also integration over the states of the continuous spectrum:

$$\Delta\psi_n^{(1)}(\vec{r}) = \sum_{k\neq n} \frac{V_{nk}\psi_k^{(0)}(\vec{r})}{E_n^{(0)} - E_k^{(0)}} + \int \frac{d\vec{p}V_{pk}\psi_p^{(0)}(\vec{r})}{(2\pi)^3(E_n^{(0)} + p^2/2\mu}. \tag{34}$$

It was precisely this latter (integration over the continuous spectrum) that was omitted in the calculations in [17, 18].

Let us estimate the contribution of the term in formula (24), which was not taken into account in [17, 18] and which corresponds to integration over the continuous spectrum, following mainly [23].

In accordance with [23],

$$\Delta\psi_{1s}^{\text{непр}}(\vec{r}) = -\int \frac{V_{1s,\vec{p}}\psi_{\vec{p}}^{(0)}(\vec{r})d^3p}{(2\pi)^3[\mu\alpha^2/2 + p^2/2\mu]}, \tag{35}$$

$$V_{1s,\vec{p}} = \int \psi_{1s}^{(0)}(\vec{r}')V_s(r')\,\psi_{\vec{p}}^{(0)}(\vec{r})d^3r',$$

$$\psi_{\vec{p}}^{(0)}(\vec{r}) = e^{\frac{\pi}{2}\xi}\Gamma(1 - i\xi)\cdot e^{i\vec{p}\vec{r}}\cdot F(i\xi; \vec{p}\vec{r} - pr),$$

$$\xi = \frac{\mu\alpha}{p}.$$

Since the main contribution to this integral comes from the values $p \gg \mu\alpha$, we can approximately set $\xi = 0$ in the expression for the wave functions of the continuous spectrum, as a result of which they are transformed into ordinary plane waves.

Using simple calculations, we obtain in this approximation

$$\Delta\psi_{1s}(\vec{r}) = -\frac{1}{4\pi}\int U_s(r')\,\psi_{1s}^{(0)}(\vec{r}')\frac{e^{-\mu\alpha|\vec{r}-\vec{r}'|}}{|\vec{r}-\vec{r}'|}d^3r', \tag{36}$$

$$U_s = 2\mu V_s.$$

Hence, for the value of Δφ1Δ(φ) we have:

$$\Delta\psi_{1s}(0) = \Delta\psi_{1s}(\vec{r})\,|_{r\to 0} = -\int U_s(r')\,\psi_{1s}^{(0)}(\vec{r}')\frac{e^{-\mu\alpha r'}}{r'}r'^2dr' \tag{37}$$

$$= -\psi_{1s}^{(0)}(0)\int_0^\infty \frac{U_s(r')r'^2dr'}{r'} = \psi_{1s}^{(0)}(0)\alpha_s < r^{-1} >_s \approx \psi_{1s}^{(0)}(0)\frac{a_s}{r_s}.$$

Here $a_s = \int U_s(r')r'^2dr'$ is the $\pi^+\pi^-$- scattering length, and $< r >_s^{-1} \approx 1/r_s$ is the average value of the reciprocal distance over the region of action of the strong interaction. Thus, if the contribution of the discrete spectrum to the relative corrections to the ground state wave functions is negligible

$$\frac{\Delta\psi_{1s}^{\text{дискр}}(0)}{\psi_{1s}^{(0)}(0)} \sim \frac{a_s}{r_B} \sim 10^{-3}\,, \tag{38}$$

then the contribution of the continuous spectrum

$$\frac{\Delta\psi_{1s}^{\text{непр}}(0)}{\psi_{1s}^{(0)}(0)} \sim \frac{a_s}{r_s} \sim 1 \tag{39}$$

can reach 100%.

Since the exact expression for the strong interaction potential is unknown, the result obtained means that the uncertainty in our knowledge of the wave functions of hadronic atoms at short distances is practically 100%.

At first glance, the relative probabilities of the formation of hadronic atoms in various *ns*-states, determined primarily by the behavior of the wave functions of these atoms at short distances, cannot be calculated with an accuracy of better than 1%.

However, if, along with the corrections to the values of the wave functions of the ground *1s*-state, we also consider the corrections to the wave functions of *ns*-states within the approximation used (the plane-wave approximation for continuous-spectrum wave functions), then it is easy to verify that they have the form

$$\Delta\psi_{ns}(\vec{r}) = -\frac{1}{4\pi}\int \psi_{ns}^{(0)}(\vec{r}')\frac{e^{-\frac{\mu\alpha|\vec{r}-\vec{r}'|}{n}}}{|\vec{r}-\vec{r}'|}U_s(r')d^3r' \tag{40}$$

$$= -\psi_{ns}^{(0)}(0)\int \frac{U_s(r')}{|\vec{r}-\vec{r}'|}d^3r'\; O\left(\frac{r_s}{r_B}\right).$$

It follows directly from this that the relation

$$c_n = \frac{\Delta\psi_{ns}(0)}{\psi_{ns}^{(0)}(0)} = \int U_s(r')r'dr' + O(10^{-3}) \tag{41}$$

with an accuracy of up to values of the order of $a_s/r_B \sim 10^{-3}$ is independent of the value of the principal quantum number n.

Then, from (10) for the case of the point source we obtain

$$\frac{\mathfrak{w}_{n_1 s}}{\mathfrak{w}_{n_2 s}} = \left|\frac{\psi_{n_1 s}(0)}{\psi_{n_2 s}(0)}\right|^2 = \left|\frac{\psi_{n_1 s}^{(0)}(0)(1+c_{n_1})}{\psi_{n_2 s}^{(0)}(0)(1+c_{n_2})}\right|^2$$

$$\left|\frac{\psi_{n_1 s}^{(0)}(0)}{\psi_{n_2 s}^{(0)}(0)}\right|^2 + O\left(\frac{a_s}{r_B}\right) = \left(\frac{n_2}{n_1}\right)^3 = O(10^{-3})\,. \tag{42}$$

Accounting for the effects of finite source sizes leads, as in [24], to corrections of the order of $< r^2 >_p / r_B^2$ to this result.

Finally, for the purposes of interpreting the results of the DIRAC experiment, the ratios $\mathfrak{w}_{ps}/\mathfrak{w}_{ns}$ at $p \sim \mu\alpha,$ are also important, where $\mathfrak{w}_{ps}$ is the probability of the formation of free $\pi^+\pi^-$-pairs in a state with zero orbital angular momentum and a relative momentum value equal to $p$.

The corrections to the wave functions of the continuous spectrum obtained in [24] in the same approximations have the form

$$\Delta\psi_{ps}(0) = -\int_0^\infty U_s(r')\,\psi_{ps}^{(0)}(\vec{r}')\frac{e^{ipr'}}{r'}r'^2dr' \tag{43}$$

For $p \sim \mu\alpha \ll m_\rho$, we get

$$\Delta\psi_{ps}(0) = -\psi_{ps}^{(0)}(0)\int\limits_0^\infty U_s(r')r'dr' = \psi_{ps}^{(0)}(0)\frac{\Delta\psi_{ns}(0)}{\psi_{ns}^{(0)}(0)} + O\left(\frac{p}{m_p}\right). \tag{44}$$

Therefore, in these approximations, the relation

$$\frac{\varpi_{ps}}{\varpi_{ns}} = \left|\frac{\psi_{ps}^{(0)}(0)}{\psi_{ns}^{(0)}(0)}\right|^2, \tag{45}$$

obtained in [24] remains valid with an accuracy of about $10^{-3}$.

However, such simple expressions (40), (41) for the corrections to the wave functions of *ns*-states were obtained by using the plane-wave approximation for the wave functions of the continuous spectrum, the accuracy of which requires further study. As shown below, this approximation ignores contributions of the order of $(a_s/r_B)\ln(r_B/r_s) \sim 10^{-2}$. To take them into account within the Rayleigh–Schrödinger perturbation theory discussed above, it would be necessary to perform successive calculations of expressions (35) with the Coulomb wave functions of the continuous spectrum, which is associated with rather cumbersome calculations. A simpler and more elegant solution to the problem under discussion is possible within a variant of perturbation theory proposed by Ya. B. Zeldovich [22], which allows corrections to the wave function of an arbitrary discrete state to be expressed in terms of the unperturbed wave function of that same state, without resorting to expansion (24).

A brief summary of the main ideas of Ya. B. Zeldovich's method and its application to assessing the influence of strong interactions on the behavior of the wave functions of elementary hadronic atoms at short distances constitutes the content of the next section of the article.

### 2.2 Application of Ya. B. Zeldovich's perturbation theory to assessing strong interaction effects

The popularity of perturbation theory in the Rayleigh–Schrödinger form is explained by its universality, i.e., its applicability to the description of perturbations of any quantum-mechanical systems, from the simplest one-dimensional (linear (an)harmonic oscillator) to infinite-dimensional (quantized fields). However, this universality comes at the cost of rather cumbersome calculations.

When considering one-dimensional potential problems of quantum mechanics, it is sometimes convenient to apply an alternative perturbation theory method proposed by Ya. B. Zeldovich [22]. This method is based on well-known results in the theory of second-order ordinary differential equations [25].

Let us demonstrate the advantages of this method using the example of our problem of the influence of the strong π+π-interaction on the behavior of the wave functions of $A_{2\pi}$-atoms at short distances. The strong interaction potential $V_s(r)$ is spherically symmetric:

$$V_s(r) = V_s(|r|) \equiv V_s(r), \qquad r = |\vec{r}|.$$

In this case, as is known [24], the variables in the Schrödinger equation (8) are separated, and it reduces to the equation for the radial wave functions $R_{nl}(r)$:

$$-\frac{1}{r}\frac{d^2}{dr^2}rR_{nl}(r) + U(r)R_{nl}(r) = -\kappa_{nl}^2 R_{nl}(r), \tag{46}$$

$$U(r) = 2\mu V(r) + \frac{l(l+1)}{r^2}, \qquad V(r) = V_c(r) + V_s(r),$$

$$\psi(\vec{r}) = \psi_{nlm}(\vec{r}) = Y_{lm}(\Theta, \Phi)R_l^n(R),$$

$$\int_0^\infty R_{nl}^2(r)r^2dr = \int |\psi_{nlm}(\vec{r})|\, d^3r = 1$$

$$\kappa_{nl}^2 = -2\mu E\,, \quad E < 0\,.$$

here $Y_{lm}(\Theta,\Phi)$ are the usual spherical harmonics [26], and $\Theta$, $\Phi$ are the polar and azimuthal angles determined from the relations

$$x = r\cdot\sin\Theta\cdot\cos\Phi\,, \quad y = r\cdot\sin\Theta\cdot\sin\Phi\,, \quad z = r\cdot\cos\Theta\,.$$

Introducing the reduced radial wave functions $\chi_{nl}(r) = rR_{nl}(r)$, normalized by the condition

$$\int_0^\infty |\chi_{nl}(r)|^2 dr = 1\,,$$

we rewrite equation (46) as

$$-\chi_{nl}''(r) + [U_0(r) + U_s(r)]\chi_{nl}(r) + \kappa_{nl}^2\chi_{nl}(r) = 0\,,$$

$$U_0(r) = 2\mu U_c(r) + \frac{l(l+1)}{r^2}\,, \tag{47}$$

$$\chi_{nl}''(r) \equiv \frac{d^2}{dr^2}\chi_{nl}(r)\,.$$

We denote the solutions of the unperturbed (Coulomb) problem by $\chi_{nl}^{(0)}(r)$, and the corresponding eigenvalues by $\kappa_{nl}^{(0)^2}$, so that

$$\chi_{nl}^{(0)}(r) + U_0(r)\chi_{nl}^{(0)}(r) + \kappa_{nl}^{(0)^2}\chi_{nl}(r) = 0\,. \tag{48}$$

The strong interaction changes both the wave functions $\chi_{nl}(r)$ and the eigenvalues $\kappa_{nl}^2$.

Introducing the notation

$$\Delta\chi_{nl}(r) = \chi_{nl}(r) - \chi_{nl}^{(0)}(r), \quad \Delta\kappa_{nl}^2 = \kappa_{nl}^2 - (\kappa_{nl}^2)^2\,,$$

for the quantity $\Delta\chi_{nl}(r)$, we obtain the equations

$$-(\Delta\chi_{nl})''(r) - \varpi_{nl}^{(0)}(r)\Delta\chi_{nl}(r) = \varpi_{nl}(r)\left(\chi_{nl}^{(0)} + \Delta\chi_{nl}(r)\right), \tag{49}$$

where

$$\varpi_{nl}(r) = U_s(r) + \Delta\kappa_{nl}^2\,, \quad \varpi_{nl}^{(0)}(r) = U_0(r) + \kappa_{nl}^2\,.$$

Expanding the quantities $\Delta\chi_{nl}$ and $\Delta\kappa_{nl}^2$ in a series in powers of perturbation

$$\Delta\chi_{nl} = \sum_{k=1}^{\infty}(\Delta\chi_{nl})^{(k)}, \quad (\Delta\chi_{nl})^{(k)} \sim (U_s)^k\,,$$

$$\Delta\kappa_{nl}^2 = \textstyle\sum_{k=1}^{\infty}(\Delta\kappa_{nl}^2)^{(k)}, \quad (\Delta\kappa_{nl}^2)^{(k)} \sim (U_s)^k,$$

we obtain in the first order of perturbation theory

$$-(\Delta\chi_{nl})^{(1)''}(r) - \varpi_{nl}^{(0)}(r)\left(\Delta\chi_{nl}(r)\right)^{(1)} = \varpi_{nl}^{(1)}(r)\chi_{nl}^{(0)}\,, \tag{50}$$

$$\varpi_{nl}^{(1)}(r) = U_s(r) + (\Delta\kappa_{nl}^2)^{(1)},$$

where

$$(\Delta\kappa_{nl}^2)^{(1)} = -\int_0^\infty U_s(r)\left(\kappa_{nl}^{(0)}\right)^2 dr\,, \tag{51}$$

according to [26], so that

$$\int_0^\infty \mathfrak{w}_{nl}^{(1)}(r)\left(\chi_{nl}^{(0)}\right)^2 dr = 0\,. \tag{52}$$

From the theory of ordinary differential equations [25], it is known that the solution of the inhomogeneous equation (50) can be expressed in terms of two linearly independent solutions $\left(\chi_{nl}^{(0)}, \phi_{nl}^{(0)}\right)$ of the corresponding homogeneous equation (48).

Their Wronskian

$$W(r) = \chi_{nl}^{(0)}(r)\left(\phi_{nl}^{(0)}(r)\right)' - \left(\chi_{nl}^{(0)}(r)\right)'\phi_{nl}(r)^{(0)},$$

by virtue of (47) satisfying the equation $W(r)' = 0$, is a constant value $W(r) = =$ const, which can be set equal to unity without loss of generality.

Consequently, the solution of the inhomogeneous equation (50) can be represented as

$$\Delta\chi_{nl}^{(1)}(r) = C\chi_{nl}^{(0)}(r) - \phi_{nl}^{(0)}(r)\int_0^\tau dr'\,\mathfrak{w}_{nl}^{(1)}(r')\left(\chi_{nl}^{(0)}(r')\right)^2 \tag{53}$$

$$+\chi_{nl}^{(0)}(r)\int_0^r dr'\,\mathfrak{w}_{nl}^{(1)}(r')\chi_{nl}^{(0)}(r')\phi_{nl}^{(0)}(r')\,.$$

Here *C* is an arbitrary constant, the value of which is fixed by the condition

$$\int_0^\infty \chi_{nl}^{(0)}(r)\Delta\chi_{nl}^{(1)}(r)dr = 0\,. \tag{54}$$

The first-order correction to the wave function must be orthogonal to the unperturbed wave function [26].

In [22], it is proved that

$$\left(\frac{\phi_{nl}^{(0)}(r)}{\chi_{nl}^{(0)}(r)}\right)' = -\frac{1}{\left(\chi_{nl}^{(0)}(r)\right)^2}\,, \tag{55}$$

whence

$$\phi_{nl}^{(0)}(r) = -\chi_{nl}^{(0)}(r)\int_{r_0}^r \frac{dr'}{\left(\chi_{nl}^{(0)}(r')\right)^2}\,. \tag{56}$$

Substituting (56) into (53), as a result of simple calculations we obtain

$$\Delta\chi_{nl}^{(1)}(r) = \chi_{nl}^{(0)}(r)\left\{C + \int_0^r \frac{dr''}{\left(\chi_{nl}^{(0)}(r'')\right)^2}\int_0^{r''} dr'\,\mathfrak{w}_{nl}^{(1)}(r')\left(\chi_{nl}^{(0)}(r')\right)^2\right\}\,. \tag{57}$$

Thus, the result for the correction $\Delta\chi_{nl}^{(1)}(r)$ to the wave function of an arbitrary discrete state is expressed in terms of the unperturbed wave function of this state and, in contrast to Rayleigh-Schrödinger perturbation theory, does not require knowledge of the entire spectrum of the unperturbed problem.

The value of the constant *C* is fixed by condition (54) as follows

$$C = -\int_0^\infty dr \left(\chi_{nl}^{(0)}(r')\right)^2 \int_0^r \frac{dr''}{\left(\chi_{nl}^{(0)}(r'')\right)^2} \int_0^{r''} dr'\, \mathfrak{w}_{nl}^{(1)}(r') \left(\chi_{nl}^{(0)}(r')\right)^2 . \qquad (58)$$

Relations (51), (57), (58) form the basis of Ya. B. Zeldovich's perturbation theory [22].

In the following, we discuss corrections only to the values of the wave functions of *ns*-states $|ns\rangle \equiv |n00\rangle$.

Since, according to (57), the correction to the value of the wave function of any state is proportional to the value of the unperturbed wave function, it is convenient to introduce the relative correction value

$$R_{n0}(r) = \frac{\Delta\psi_{n0}(r)}{\psi_{n0}^{(0)}(r)} = \frac{\Delta\chi_{n0}^{(1)}(r)}{\chi_{n0}^{(0)}(r)} = C + \int_0^r \left(\frac{dr''}{\chi_{nl}^{(0)}(r'')}\right)^2 \int_0^{r''} dr'\, \mathfrak{w}_{n0}^{(1)}(r') \left(\chi_{n0}^{(0)}(r')\right)^2 , \qquad (59)$$

where $C$ is given by relation (58).

This value reaches its maximum, as can be seen below, at $r = 0$

$$\max R_{n0}(r) = R_{n0}(0) = C \qquad (60)$$

$$= -\left\{ \int_0^\infty dr |\chi_{nl}^{(0)}(r)|^2 \int_0^\infty dr \left(\chi_{nl}^{(0)}(r')\right)^2 \int_{r_0}^r \left(\frac{dr''}{\chi_{nl}^{(0)}(r'')}\right)^2 \int_0^{r''} dr'\, \mathfrak{w}_{nl}^{(1)}(r') \left(\chi_{nl}^{(0)}(r')\right)^2 \right\} .$$

The calculations lead to the following result

$$R_{n0}(0) = \frac{\Delta\psi_{n0}(0)}{\psi_{n0}^{(0)}(0)} = -\int_0^\infty r^2 e^{-\frac{2v}{n}r} U_s(r) \left[\frac{1}{r} - 2v\ln(vr) + P_{2n}(vr)\right] dr , \qquad (61)$$

where $P_{2n}(t)$ is the polynomial of degree *2n*.

It is easy to see that the contribution to the ratio $\Delta\psi_{n0}(0)/\psi_{n0}^{(0)}(0)$ that depends on *n* is of the order of $a_s/r_B \sim 10^{-3}$. Contributions of the order of $a_s/r_s$ and $(a_s/r_B)\ln(r_B/a_s)$ do not depend on *n*.

This result is a consequence of the relation

$$\frac{|d\psi_{n0}(r)/dr|_{r=0}}{\psi_{n0}^{(0)}(0)\big|_{r=0}} = -v = -\mu\alpha = \mathrm{const}(n) , \qquad (62)$$

already used by us earlier (see eq. (18)).

Finally, we present the expression (63) for $R_{n0}(r)$, valid for the values $r \ll r_B$

$$R_{n0}(r) = -\int_r^\infty r^2 U_s(r) \left[\frac{1}{r} - 2v\ln vr\right] dr + \left[\frac{1}{r} - 2v\ln vr\right] \int_0^r r^2 U_s(r) dr + O\left(\frac{r_s}{r_B}\right) . \qquad (63)$$

From (63) it is clear that taking into account the effects of strong $\pi^+\pi^-$-interactions in first-order perturbation theory reduces to multiplying the unperturbed wave functions of pionium atoms by the functions $R_n(r)$, which are practically independent of *n*:

$$\psi_{n0}(r) = R_n(r)\psi_{n0}^{(0)}(r) , \qquad (64)$$
$$R_n(r) = 1 + R_{n0}(r) ,$$

and this relation is valid for any form of the potential $U_s$.

The independence of the $R_n(r)$ functions from the value of *n* can be explained as follows.

Let us write explicit expressions (65) for the first few reduced Coulomb radial functions $\chi_{n0}(\rho)$, $\rho = r/r_B$, multiplied by $n^{3/2}$ for convenience

$$\chi_{10}(\rho) = 2\rho\exp(-\rho) = 2\left(\rho - \rho^2 + \frac{\rho^3}{2}\right) + O(\rho^4) , \qquad (65)$$

$$2\sqrt{2}\chi_{20}(\rho) = 2\rho\left(1 - \frac{\rho}{2}\right)\exp\left(-\frac{\rho}{2}\right) = 2\left(\rho - \rho^2 + \frac{\rho^3}{8}\right) + O(\rho^4) ,$$

$$3\sqrt{3}\chi_{30}(\rho) = 2\rho\left(1-\frac{2\rho}{3}+\frac{2\rho^2}{27}\right)\exp\left(-\frac{\rho}{3}\right) = 2\left(\rho-\rho^2+\frac{19\rho^3}{54}\right)+O(\rho^4),$$

$$4\sqrt{4}\chi_{40}(\rho) = 2\rho\left(1-\frac{3\rho}{4}+\frac{\rho^2}{8}-\frac{\rho^3}{192}\right)\exp\left(-\frac{\rho}{4}\right) = 2\left(\rho-\rho^2+\frac{11\rho^3}{32}\right)+O(\rho^4).$$

It is easy to see that the coefficients of the first two terms of the expansion coincide for all values of $n$ and the difference begins only at values of $O(10^{-3})$. It is precisely this coincidence that ensures that the combinations $U_s(r)r^2[1/r - 2v\ln vr]$ under the sign of the integral determining the value of $R_n(p)$ (63) differ from each other by $O(r_s/r_B{\sim}10^{-3})$. If it were absent, the accuracy could be an order of magnitude worse, i.e. $O(10^{-2}) = (r_s/r_B)\ln(r_B/a_s)$. The stated independence of the functions $R_n(p)$ from the value of $n$ is accurately confirmed by numerical calculations [27].

Numerical calculations of the relative correction of the wave function at zero in all orders of perturbation theory for the first four values of the principal quantum number and the Yukawa-type potential $U_s$ lead to its value $R_{n0}(0) = R_0(10) + O(10^{-3}){\sim}0.77$. In first-order perturbation theory $R_{10}^{(1)}(0){\sim}0.55$ and for an exponentially decreasing potential, they result in the value $R_{n0}(0){\sim}0.40$ ($R_{10}^{(1)}(0){\sim}0.28$) [27].

The high sensitivity of the correction value $R_{n0}(0)$ to the type of the potential $U_s$ makes it desirable (and, in the case of measurement, possible) to experimentally determine its value. One of the most important consequences of (64) is the fact that the law $\mathfrak{w}_n{\sim}n^{-3}$ for the probabilities of the formation of dimesoatoms in *ns*-states, derived under the assumption that their wave functions are purely Coulomb, remains valid in the general case. Indeed, substituting (64) into (15) and replacing $M(\vec{r})$ by $\widetilde{M}(\vec{r}) = M(\vec{r})R(r)$, we obtain the following equation

$$\mathfrak{w}_n{\sim}\left|\iint \widetilde{M}(\vec{r})\,\psi_n^{(0)}dr\right|^2{\sim}\left|\psi_n^{(0)}(0)\right|^2\left|\iint \widetilde{M}(\vec{r})\,dr\right|^2{\sim}n^{-3}.$$

Thus, the ratio of the probabilities of the formation of pionium atoms in different *ns*-states turns out to depend only on the behavior of the unperturbed (Coulomb) wave functions of pionium at short distances and the ratio between the sizes of the formation region of $\pi^+\pi^-$-pairs and the sizes of $\pi^+\pi^-$-atoms. Due to the specific properties of the Coulomb wave functions of pionium atoms $\psi_{ns}^{(c)}(0) \equiv \psi_{n0}^{(0)}(0)$ in *ns*-states, reflected by relations (18), (65), these relations, with an accuracy of about $r_s/r_B{\sim}10^{-3}$, turn out to be independent of the insufficiently studied details of the dynamics of the strong interaction and are equal to

$$\frac{\mathfrak{w}_{n_{10}}}{\mathfrak{w}_{n_{20}}} = \left(\frac{n_2}{n_1}\right)^3 + O(10^{-3}). \qquad (66)$$

**3 Deser's formula for the level shift in terms of the effective radius of the strong interaction and the correction to the Coulomb wave function at the origin**

The results obtained significantly simplify the task of theoretical interpretation of the DIRAC experiment [9], for which only the values of the probability ratios $\mathfrak{w}_{n_{10}}/\mathfrak{w}_{n_{20}}$ for the formation of pionium atoms in various *ns*-states are important.

However, if it were possible to measure the absolute value of the probability of the formation of pionium atoms in the ground state, the possibility of which was indicated in [16–18], then, together with measuring the energy difference between the levels of the *2s*- and *2p*- states [9, 16, 28]

$$\Delta E_{2s-2p} = \Delta E_{2s-2p}^{s} + \Delta E_{2s-2p}^{vac}, \qquad (67)$$

this could allow estimating the effective radius of the strong $\pi^+\pi^-$-interaction $r_s$ as $r_s{\sim}a_s/c_1$, which, along with the scattering length $a_s$, is an important characteristic of $\pi^+\pi^-$-scattering.

Let us demonstrate this. Returning to first-order Rayleigh–Schrödinger perturbation theory, we write the expression for the shift of the *ns*-level of a hadronic atom due to the strong interaction in the form

$$\Delta E_{n0}^{s} = \int |\psi_{n0}^{c}(r)|^2 V_s(r) d^3r$$

$$\approx |\psi_{n0}^{c}(0)|^2 \int V_s(r) d^3r = 4\pi |\psi_{n0}^{c}(0)|^2 \int V_s(r) r^2 dr\,, \quad (68)$$

where

$$\int V_s(r) r^2 dr = -\frac{a_s}{2\mu}\,. \quad (69)$$

As a result, we obtain the Deser formula for the shift of the *ns*-levels of hadronic atoms, which coincides with formula (1):

$$\Delta E_{n0}^{s} = -\frac{2\pi}{\mu} a_s |\psi_{n0}^{c}(0)|^2\,. \quad (70)$$

When applied to $\pi^+\pi^-$-atoms, this formula has the form

$$\Delta E_{n0}^{s} = -\frac{2\pi}{3\mu}(2a_0 + a_2)|\psi_{n0}^{c}(0)|^2\,. \quad (71)$$

Let us estimate the last quantity. For the scattering lengths $a_0(m_\pi = 1) = 0.22$, $a_2(m_\pi = 1) = -0.044$ [28], we obtain

$$\Delta E_{10}^{s} = -\frac{4\pi}{3m_\pi}(2a_0 + a_2)|\psi_{10}^{c}(0)|^2 \approx -3.56 \text{ эВ}\,, \quad (72)$$

$$\Delta E_{2p-2s}^{s} = \Delta E_{20}^{s} = -\frac{4\pi}{3m_\pi}(2a_0 + a_2)|\psi_{20}^{c}(0)|^2 \approx -0{,}45 \text{ эВ}\,, \quad (73)$$

The possibility of measuring the magnitude of $\Delta E_{2p-2s}^{s}$ (73) is due to the fact that the contribution of the vacuum polarization $\Delta E_{2p-2s}^{vac}$ в $\Delta E_{2p-2s}$ (67) is accurately calculated within QED [29]. According to the data presented in [28], its value is $\Delta E_{2p-2s}^{vac} = -0.107$.

Now note that the relation

$$\frac{a_s}{c_n} = -\frac{\int_0^\infty U_s(r) r^2 dr}{\int_0^\infty U_s(r) r dr} = \langle r \rangle_s \quad (74)$$

allows reformulating the Deser equation (70) in terms of the average radius over the strong interaction region $\langle r \rangle_s \sim r_s \sim m_\rho^{-1}$, where $m_\rho$ is the $\rho$-meson mass, and the relative correction $c_n = \frac{\Delta\psi_{n0}(0)}{\psi_{n0}^{c}(0)} = c_1 + O(10^{-3})$ to the Coulomb wave function at zero is as follows

$$\Delta E_{n0}^{s} = -\frac{2\pi}{\mu} \langle r \rangle_s \; c_n\, |\psi_{n0}^{c}(0)|^2\,. \quad (75)$$

Note also that, unlike (70), formula (75) can be written in terms of the exact wave function (strongly modified at zero [27])

$$\Delta E_{n0}^{s} = -\frac{2\pi}{\mu} \langle r \rangle_s \; \frac{c_n}{(1 + c_n)^2}\, |\psi(0)_{n0}|^2\,. \quad (76)$$

The factorization properties of the wave function itself $\psi_{n0}(0) = \psi_{n0}^{c}(0)(1 + c_n)$ are analogous to the factorization properties of equations (1), (2), and (75).

Finally, we will write in general form the system of equations

$$(1 + c_n)^2 = a|\psi_{n0}(0)|^2\,, \quad a = \pi(n r_B)^3\,, \quad (77)$$

$$< r >_s = -b\frac{\Delta E_{k0}^s}{c_k}\,, \qquad b = \mu(kr_B)^3/2\,,$$

that enables, in particular, under the condition of measuring the quantity $|\psi_{10}(0)|^2$ or, which is practically the same, the relative correction $c_1$, to determine the average radius of the strong interaction region from the experimentally measured quantity $E_{2s-2p}$. It is similar to how the combined measurement of $\Delta E_{2s-2p}$ and $\Gamma_{10}$ leads to an accurate determination of the scattering lengths $a_0$ and $a_2$ in their combinations $a_s$, $a_{ce}$ in (1), (2).

This possibility is of interest in connection with well-known experiments on measuring the shift of the *1s*-levels of $A_{\pi^- p}$ [6], $A_{K^- p}$ [8] atoms, the energy difference between the *2s*- and *2p*-levels of $A_{2\pi}$, $A_{K^+\pi^-}$-atoms [9, 28] and the recent precise measurement of the Lamb shift in muonium [30].

**Acknowledgments**

The author is grateful to Dr. Alexander Tarasov† for discussing the results obtained. He also thanks Dr. Elena Pavlova for technical assistance in preparing this article. The article is dedicated to the 90th anniversary of Prof. Leonid Nemenov, who demonstrated in [28] the possibility of measuring the magnitude of $\Delta E_{2p-2s}^s$ (73).